\documentclass[9pt,twocolumn,floatfix,pra]{revtex4-1}
\usepackage{amsmath,amssymb,amsthm,mathrsfs,amsfonts,dsfont} 
\usepackage{amsmath,amssymb,amstext}
\usepackage{textcomp}
\usepackage{pbox}
\usepackage{tikz}
\usepackage{bm}
\usepackage{dcolumn}
\usepackage{booktabs}
\usepackage[scaled]{helvet}
\usepackage{sansmath}
\usepackage{graphicx}
\usepackage{transparent}
\usepackage{color}
\usepackage{gensymb}
\usepackage{multirow}
\usepackage{afterpage}
\usepackage{url}

\usepackage[colorlinks=true]{hyperref}
\usepackage[caption=false]{subfig}

\begin{document}

\title{Temporal multimode storage of entangled photon pairs}

\author{Alexey Tiranov$^{1}$}
\author{Peter C. Strassmann$^{1}$}
\author{Jonathan Lavoie$^{1}$}
\altaffiliation[Current address: ]{Present address: Department of Physics and Oregon Center for Optical Molecular \& Quantum Science, University of Oregon, Eugene, OR 97403, USA}
\email{jlavoie@uoregon.edu}

\author{Nicolas Brunner$^{2}$}
\author{Marcus Huber$^{1,3}$}

\author{Varun B.~Verma$^{4}$}
\author{Sae Woo Nam$^{4}$}
\author{Richard P.~Mirin$^{4}$}
\author{Adriana E.~Lita$^{4}$}
\author{Francesco Marsili$^{5}$}

\author{Mikael Afzelius$^{1}$}
\author{F\'{e}lix Bussi\`{e}res$^{1}$}
\author{Nicolas Gisin$^{1}$}

\affiliation{$^{1}$Groupe de Physique Appliqu\'ee, Universit\'e de Gen\`eve, CH-1211 Gen\`eve, Switzerland}
\affiliation{$^{2}$D\'epartement Physique Th\'eorique, Universit\'e de Gen\`eve, CH-1211 Gen\`eve, Switzerland}
\affiliation{$^{3}$Institute for Quantum Optics and Quantum Information (IQOQI), Austrian Academy of Sciences, Boltzmanngasse 3, A-1090 Vienna, Austria}
\affiliation{$^{4}$National Institute of Standards and Technology, 325 Broadway, Boulder, CO 80305, USA}
\affiliation{$^{5}$Jet Propulsion Laboratory, California Institute of Technology, 4800 Oak Grove Dr., Pasadena, California 91109, USA}


\newcommand{\ket}[1]{\vert#1\rangle}
\newcommand{\bra}[1]{\langle#1\vert}
\newcommand{\xp}[1]{\langle#1\rangle}
\newcommand{\YSO}{Y$_2$SiO$_5$}
\newcommand{\nd}[0]{Nd$^{3+}$:Y$_2$SiO$_5$}
\newcommand{\X}{\textbf{X}}
\newcommand{\Y}{\textbf{Y}}
\newcommand{\e}{\mathrm{e}}
\newcommand{\Vis}{\mathcal{V}}
\newcommand{\C}{\mathcal{C}}
\newcommand{\mathbbm}[1]{\text{\usefont{U}{bbm}{m}{n}#1}} 
\newcommand{\braket}[2]{\langle#1\vert#2\rangle}
\newcommand{\suppmat}[0]{{Appendix}}


\begin{abstract}
Multiplexed quantum memories capable of storing and processing entangled photons are essential for the development of quantum networks. In this context, we demonstrate and certify the simultaneous storage and retrieval of two entangled photons inside a solid-state quantum memory and measure a temporal multimode capacity of ten modes. This is achieved by producing two polarization entangled pairs from parametric down conversion and mapping one photon of each pair onto a rare-earth-ion doped (REID) crystal using the atomic frequency comb (AFC) protocol. We develop a concept of indirect entanglement witnesses, which can be used as Schmidt number witness, and we use it to experimentally certify the presence of more than one entangled pair retrieved from the quantum memory. Our work puts forward REID-AFC as a platform compatible with temporal multiplexing of several entangled photon pairs along with a new entanglement certification method useful for the characterisation of multiplexed quantum memories.
\end{abstract}

\maketitle 

Quantum memories are key elements for developing future quantum networks~\cite{Kimble2008}. Optical quantum memories~\cite{Lvovsky2009,Bussieres2013} allow storage of parts of optical quantum states, for instance the storage of one photon out of a pair of entangled photons. This ability can be used to herald entanglement between stored excitations of remote quantum memories~\cite{Chou2005,Moehring2007,Yuan2008}, which is a basic resource for long-distance quantum networks. A prominent example is the quantum repeater~\cite{Briegel1998,Duan2001}, which in principle can distribute quantum entanglement over continental distances, thereby allowing long-distance quantum key distribution over scales impossible by current technologies~\cite{Korzh2015}. 

Most quantum repeater schemes require efficient multiplexing in order to achieve any useful rate of entanglement distribution~\cite{Sangouard2011}, which in turn requires quantum memories (QM) that are highly multimode~\cite{Simon2007}. Quantum memories based on ensembles of atoms provide such a resource, where different degrees of freedom can be used to achieve multimode storage, such as spatial~\cite{Lan2009,Zhou2015,Parigi2015,Dai2012,Wu2016}, spectral~\cite{Sinclair2014,Saglamyurek2016} or temporal modes~\cite{Nunn2008}. The ensemble approach also provides strong collective light-matter coupling~\cite{Gorshkov2007,Hammerer2010}, making high memory efficiencies possible. 

\begin{figure}[htp]
\centering
\includegraphics[width=0.45\textwidth]{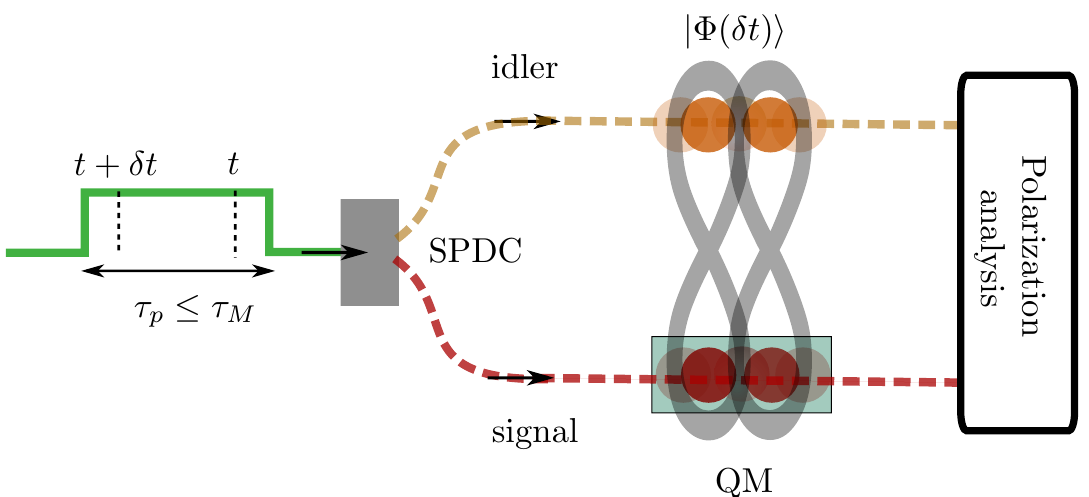}
\caption{(color online) Conceptual scheme of the experiment. Temporally multiplexed photon pairs generated from spontaneous parametric down conversion (SPDC) are stored in a multimode quantum memory (QM) and released after a predetermined time $\tau_M$. The pump is pulsed with a 10~MHz repetition rate and with duration $\tau_p$. Two polarization-entangled pairs are generated probabilistically by the same pulse and separated by a time $\delta t\leq \tau_M$. The duration of the pulse equals the storage time such that the stored photons are both in the QM for a time ($\tau_M-\delta t$). To certify entanglement after absorption and remission by the QM we analyze the correlations in polarization of the four-photon state.}
\label{fig:scheme}
\end{figure}

In this work we address the challenge of demonstrating and certifying simultaneous storage of several quantum excitations in different temporal modes of a QM. We focus on temporal multimode storage in a single spatial mode, which is compatible with optical fiber technologies and therefore attractive for long-distance quantum networks. We use the atomic frequency comb (AFC) approach~\cite{Afzelius2009a}, which can achieve multimode storage for much lower optical depths compared to other ensemble-based storage techniques~\cite{Nunn2008}. This protocol is specifically developed for rare-earth-ion doped (REID) crystals~\cite{Tittel2010b}. Previous studies have demonstrated temporal highly multimode storage using the AFC scheme, but these experiments have employed either strong~\cite{Bonarota2011a,Gundogan2013,Jobez2015b} or attenuated laser pulses~\cite{Usmani2010,Jobez2015,Gundogan2015}, and true single-photon pulses~\cite{Tang2015} but without entanglement. Here we demonstrate simultaneous storage of two entangled photon pairs. The certification of multi-pair entanglement in our experiment requires novel tools, given the limited available data. This is achieved by constructing indirect entanglement witnesses, which we use experimentally to certify the presence of more than one entangled pair retrieved from the quantum memory. 

The setup of our experiment is illustrated in Fig.~\ref{fig:scheme}. Two independent pairs of entangled signal and idler photons are generated via spontaneous parametric down conversion (SPDC) within a time window  shorter than the memory time. The idler photons are at the telecommunication wavelength of 1338~nm, while the signal photons are at 883~nm~\cite{Clausen2014a}. The two signal photons are stored in the REID crystal in the same spatial mode, but in two independent temporal modes that differ by up to ten times their coherence time. After a pre-determined storage time, the two photons are re-emitted from the memory and detected by the single-photon detectors (and similarly for the idler photons).

The full characterization of the state of both photon pairs is constrained by the fact that their creation time is much smaller than the dead time of the detectors. This is typical of current single-photon detectors and it can complicate the analysis of temporally multiplexed quantum memories storing short (broadband) single photons. One obvious solution is to double the number of analyzers (and detectors), or use complex multiplexing schemes in space or frequency~\cite{Collins2013a,Donohue2014}. Here, instead, we want to use a pair of detectors on each side and apply the same projective measurement needed to analyze a single pair. This leads to a limited set of measurements and outcomes. 
Previous efforts have been devoted to addressing non-linear functions of density matrix elements, which due to the lack of convexity proved very challenging~\cite{Krenn2013, Krenn2014} and still lack assumption-free certification methods. To address this, we develop a new concept of induced witness operators, where an incomplete set of count rates can conclusively certify entanglement, or even Schmidt numbers, without any assumptions about the state. We then apply this new concept to certify the presence of two entangled pairs in our experiment.

\begin{figure}[!t]
\centering
\includegraphics[width=0.48\textwidth]{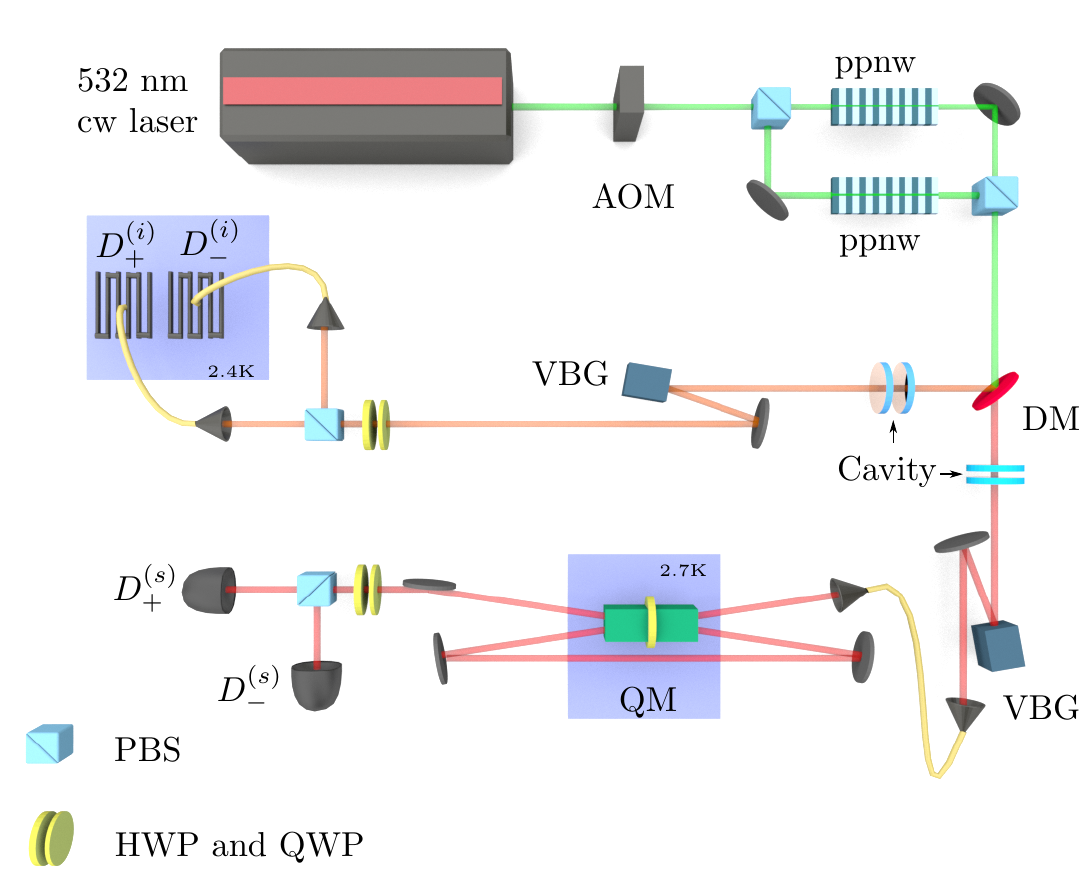}
\caption{(color online) Detailed experimental setup. Our source of polarization-entangled photon pairs is based on parametric down conversion inside two periodically poled nonlinear waveguides (ppnw) phase matched for Type-0 downconversion. Both crystals are coherently pumped by a 532~nm continuous wave (cw) laser.  The pump laser is intensity modulated by an acousto-optic modulator (AOM) at a 10~MHz repetition rate producing a train of 50~ns square pulses. Signal (883~nm) and idler (1338~nm) photons are spatially separated by a dichroic mirror (DM) and spectrally filtered using a cavity and a volume Bragg grating (VBG) in each output mode. The signal mode is coupled to a polarization-preserving solid state quantum memory (QM) based on two \nd{} crystals separated by a half-wave plate. To increase the storage efficiency, we use a double-pass configuration through the QM. Finally, we analyze the polarization states of the photon pairs with a set of half-wave (HWP) and quater-wave (QWP) plates, polarizing beamsplitter (PBS) and single photon detectors $D_\pm^{(s,i)}$; fiber-coupled superconducting nanowires on the idler side and free-space APDs on the signal side.
}
\label{fig:exp}
\end{figure}

First, we demonstrate the capability to generate two independent entangled photon pairs for further quantum storage. For this we generate polarization-entangled photon pairs from SPDC inside two nonlinear waveguides. The continuous pump laser has a central wavelength of 532~nm and is modulated in intensity to obtain a 10~MHz train of 50~ns square pulses (Fig.~\ref{fig:scheme}). This modulation defines a temporal window, corresponding to the storage time, inside which two pairs can be generated. The configuration of the nonlinear waveguides, shown in Fig.~\ref{fig:exp}, is such that photons are created in a coherent superposition of $\ket{HH}$, from the first waveguide, or $\ket{VV}$, from the second~\cite{Clausen2014a}. We approximate each pair by the state
\begin{equation}
\ket{\phi(t)} = \frac{1}{\sqrt{2}}\left( \ket{H_s, H_i}_t + \ket{V_s, V_i}_t\right),
\label{eq:psi}
\end{equation}
where $t$ is the photon pair creation time within the square pump window, $s$ and $i$ subscripts label signal and idler modes, while $\ket{H}$ and $\ket{V}$ designate horizontal and vertical polarization states of a single photon, respectively. 

Two independent polarization-entangled pairs can be generated from the same pulse, in condition that the delay between the pairs, $\delta t$, is sufficiently larger than the coherence time of one pair~\cite{deRiedmatten2004}. In this case, the joint state of two pairs is described by 
\begin{equation}
\ket{\Phi(\delta t)}=\ket{\phi(t)} \otimes \ket{\phi(t+\delta t)}.
\label{eq:2phi}
\end{equation}
The measured coherence time of a photon pair, $\tau_c=1.9$~ns, is defined by the filtering system which is applied to the both photons (for the details see~\cite{Clausen2014a}). Overall, the rate of two-fold coincident detections after storage is 200~Hz for an average pump power of 4~mW (2~mW at the input of each waveguide).

The signal mode of each pair is coupled to the QM. The latter consists of two \nd{} crystals mounted around a half-wave plate, together enabling high-fidelity polarization storage~\cite{Clausen2012, Zhou2012,Tiranov2015a}. The absorption profile of the broad resonant frequency transition of the atomic ensemble is tailored in a frequency comb using optical pumping techniques~\cite{Tiranov2015a}.  
The prepared AFC fixes the storage time to $\tau_M = 50$~ns~\cite{Afzelius2009a} and the measured total memory efficiency of the single photon is $\eta = 7(1)$\% in the double-pass configuration depicted in Fig.~\ref{fig:exp}. To analyze the correlations between the released signal and idler photons, we use a combination of quarter-wave, half-wave plates and a polarizing beam splitter (PBS) on each side. 

To detect the stored signal and the idler photons from each pair we put two single-photon detectors (SPDs) at the output ports of the PBS on each side (denoted as ``$+$'' and ``$-$'' in Fig.~\ref{fig:exp}). We use superconducting nanowire SPDs ($D_{\pm}^{(i)}$) with 75\% efficiency, 100~ns dead time and 300~ps jitter (WSi superconducting nanowire~\cite{Verma2014a}) for the idler photons. The signal photons are detected with two free-space free-running silicon avalanche photodiode ($D_{\pm}^{(s)}$) with 40\% efficiency, 1~$\mu$s dead time and 400~ps jitter.  

\begin{figure}[!t]
\centering
\includegraphics[width=0.48\textwidth]{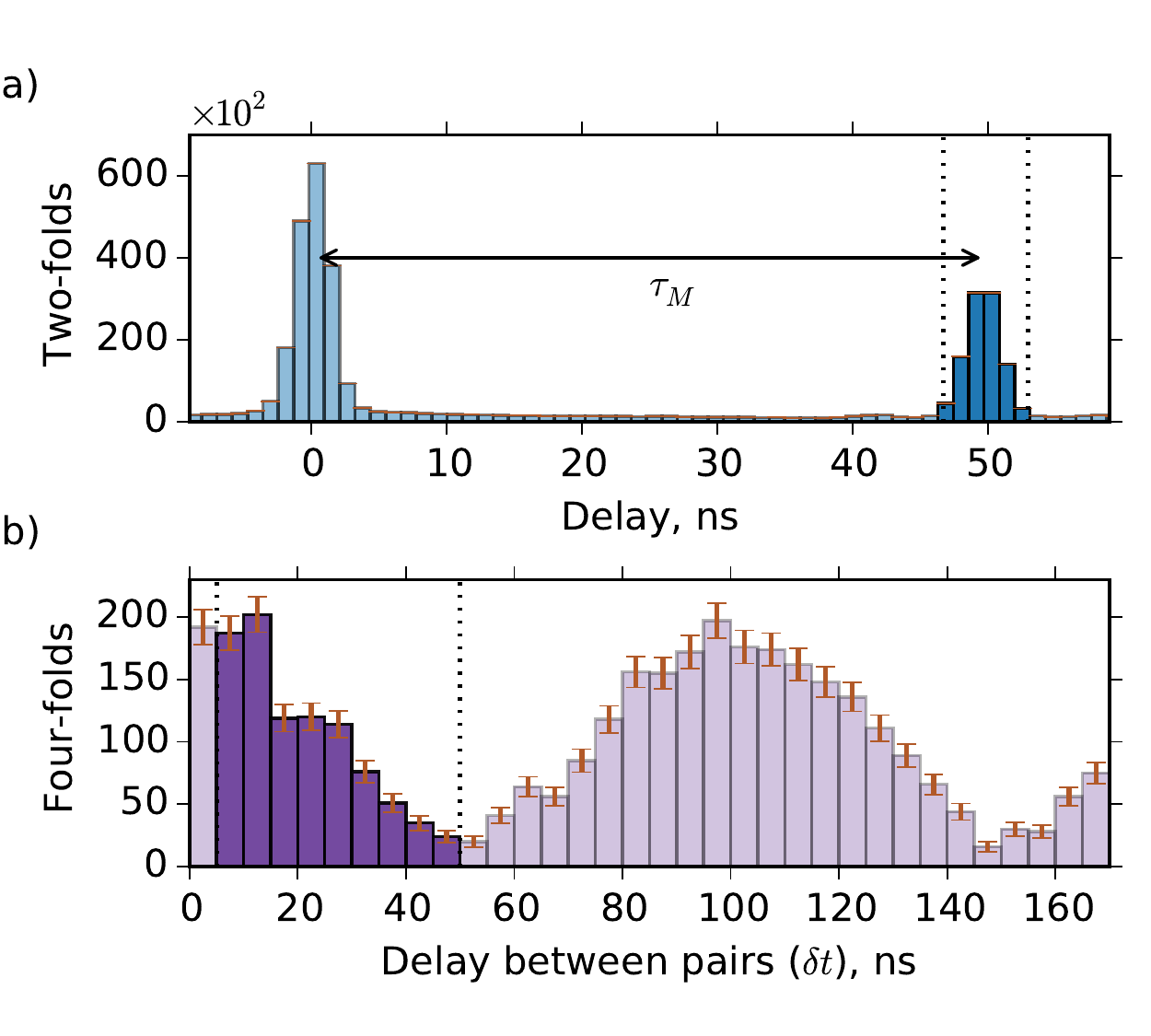}
\caption{(color online) Temporal multimode storage. (a) The two-fold coincidences between detections of the signal and idler photons as a function of the delay between two detection events. The first peak at 0 ns stems from the signal photons not absorbed by the QM while the second peak at 50~ns corresponds to the signal photons absorbed by the QM and released after the storage time. Such an histogram is accumulated for each pair of detectors between signal and idler photons. A coincidence window of 4~ns (depicted by the dotted lines) is used to calculate the rates. (b) The total four-fold coincidences collected during the experiment is plotted as a function of the delay $\delta t$ between photon pairs. 
The events corresponding to the stored state~(\ref{eq:2phi}) inside the storage time of the QM are delimited by dotted vertical lines from 5 to 50~ns.
There are 9 distinguishable time divisions, demonstrating storage of 10 modes containing single photon excitations.
For longer delay ($>50$~ns), two photons do not overlap at any time in the QM. Error bars represent one standard deviation assuming Poisson noise for the counts.}
\label{fig:delay}
\end{figure}

Figure~\ref{fig:delay}(a) shows two-fold coincidences as a function of the delay between the detections of a signal and an idler photons. The temporally resolved peak structure corresponds to the transmitted (0~ns) and stored signal photon in the QM (50~ns) from a single photon pair~(Eq.~(\ref{eq:psi})). However, to detect and analyze the four-photon state~(Eq.~(\ref{eq:2phi})) one has to look at coincidences between all four detectors (four-folds). 
Our main limitation comes from the $\sim1$~$\mu$s dead time of the signal mode detectors. Given that one photon of the first pair is detected at one output port of the PBS, the photon of the second pair cannot be detected in the same output port, as the separation is smaller than the dead time. For example, if the first photon pair is detected in $(D_+^{(s)},D_+^{(i)})$, the only way to detect the following photon pair is with the complementary detector combination $(D_-^{(s)},D_-^{(i)})$. For a given pair of measurement settings (denoting $x$ and $y$ the choice of measurement setting for Alice and Bob, respectively), there are thus four accessible event rates labelled $N_{ab,\bar{a}\bar{b}|xy}$, where $a,b = \pm$. Figure~\ref{fig:delay}(b) shows the histogram of measured four-fold events, for different delays $\delta t$ between pairs. The triangular shape is caused by the square wave pumping of the SPDC: the probability to generate two photon pairs with the delay between them equal to the pump length (in our case 50~ns) is much smaller for smaller delays.

Two detections on the idler's side, with delay $\delta t$, herald two signal photons, also with delay $\delta t$ (within their coherence times), counted and analyzed after storage in the QM. In the data analysis, the delay between two pairs is bounded by the storage time $\tau_{M}=50$~ns. Hence, we only consider the overall range delimited by two dotted vertical lines on Fig.~\ref{fig:delay}(b). Importantly, by comparing the histograms corresponding to different delays between photon pairs, we see that the presence of two excitations does not change the efficiency of the~QM. The multimode capacity of the QM is given by the number of modes (bins on the histogram) and equal to 10. Those results are direct signatures of temporal multimode capacity of our QM using simultaneous storage of two single photons.

The limited set of projection measurements used in this experiment does not allow us to probe entanglement of the four-photon state~(Eq.~(\ref{eq:2phi})) using standard tools. To certify the presence of two entangled pairs in different temporal modes we therefore devise an entanglement witness based on the event rates $N_{ab,\bar{a}\bar{b}|xy}$ for different pairs of measurement settings. 

\begin{figure}
	\centering
		\includegraphics[width=0.45\textwidth]{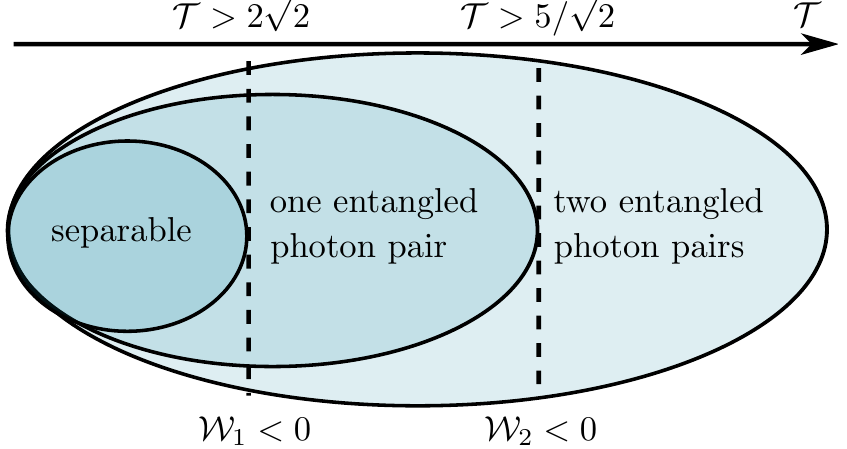}
		\caption{(color online) Entanglement witness. From the measured four-fold correlations we can compute our central figure of merit $\mathcal{T}$ (Eq.~(\ref{eq:T_4})). A value of $\mathcal{T}$ implies that the operator $\mathcal{W}_k(\mathcal{T})$ has an expectation value equal to exactly zero. For a sufficiently large value of $\mathcal{T}$ this implies that the expectation value of $W_1(\mathcal{T})$ (or $W_2(\mathcal{T})$) is negative. Since we can prove that $\mathcal{W}_1$ is an entanglement witness and $\mathcal{W}_2$ a Schmidt number witness, we can thus conclude on one or two entangled pairs.}
	\label{fig:witness}
\end{figure}

To distinguish the different temporal modes the overlap between adjacent modes should be minimized. The size of the temporal bins must be larger than the coherence time of the photon pair generated from the SPDC ($\sim2$~ns). For this reason, we consider only the events when photon pairs are separated more than 5~ns (Fig.~\ref{fig:delay}(b)), which corresponds to 2.5 times the temporal width of a photon pair. 
Specifically, we denote the final quantum state (after the memory) $\rho$, which is of dimension $4 \times 4$. Our goal is to prove that $\rho$ contains two entangled pairs (Fig.~\ref{fig:witness}) and do not have to assume \textit{a priori} that $\rho$ consists of two independent pairs (as in Eq.~(\ref{eq:2phi})). That is, violation of our entanglement witness certifies the presence of more than one entangled pair, even if the two pairs may have become correlated inside the QM. Surprisingly this is achieved even without direct access to density matrix elements, but only proportionality relations between them \suppmat.

Our witness involves two local measurement settings per party. Using Bloch vector notation, Alice's measurements are given by vectors $\hat{\sigma}_x$ (for $x=0$) and $\hat{\sigma}_y$ (for $x=1$), while Bob's measurements are $(\hat{\sigma}_x+\hat{\sigma}_y)/\sqrt{2}$ (for $y=0$) and $(\hat{\sigma}_x-\hat{\sigma}_y)/\sqrt{2}$ (for $y=1$). 
Note that this choice of measurements is typical for Bell-CHSH tests~\cite{Tiranov2015a} and suitable for the entanglement witness described below. 
For a given choice of local measurements, the event rates $N_{ab:\bar{a}\bar{b}}$ are thus proportional to $\text{Tr}(P_a \otimes P_{\bar{a}} \otimes P_b\otimes P_{\bar{b}} \rho)$, where $P_a = (\openone + a \hat{v} \cdot \hat{\vec{\sigma}} )/2$ denote qubit projectors. Our entanglement witness is based on the expression
\noindent \begin{equation}
\mathcal{T}=   \frac{1}{N}( \C_{00}+\C_{01}+\C_{10}-\C_{11})   
\label{eq:T_4}
\end{equation}
where we have defined correlation functions 
\begin{equation}
\C_{xy}=  \sum_{a,b = \pm 1} ab  \,  N_{ab,\bar{a}\bar{b}|xy}
\label{eq:e_4}
\end{equation}
and a normalization factor 
\begin{equation}
N = \frac{1}{4}\sum_{x,y=0,1} \sum_{a,b=\pm1}  N_{ab,\bar{a}\bar{b}|xy}.
\end{equation}
In the case $\rho$ contains one entangled pair (or less), the expression $\mathcal{T}$ is upper-bounded by
\begin{equation}
\mathcal{T} \leq \frac{5}{ \sqrt{2}} \simeq 3.535.
\label{eq:T_ineq}
\end{equation}
Hence any violation of the witness, i.e. $\mathcal{T} > 5 / \sqrt{2}$, implies the presence of more than one entangled pair in the output state $\rho$. For any separable state $\rho$ we prove that the expression $\mathcal{T}$ has an upper bound of $2\sqrt{2}$. This allows to use the actual value of the witness to certify the presence of entanglement in general (Fig.~\ref{fig:witness}); the value of $\mathcal{T}$ can be attributed to the expectation value of an induced operator $\mathcal{W}(\mathcal{T})$. Depending on the value of $\mathcal{T}$ the fact that this operator has a vanishing expectation value certifies entanglement of one or more entangled photon pairs.
The details of the derivation of the bounds are given in \suppmat.

\begin{table}[htp]
\centering
\renewcommand{\arraystretch}{2}
\begin{ruledtabular}
\begin{tabular}{cccc|cc}
              $\C_{00}$ & $\C_{01}$ & $\C_{10}$ & $\C_{11}$ &     $\mathcal{T}$   & $\tilde{\mathcal{T}}$  \\
\cline{1-6}
0.98(6)   &   0.92(7)   &   0.88(6)   &    -0.88(6)  & 3.67(6) &    $3.64(2)$  \\
\end{tabular}
\end{ruledtabular}
\caption{
Experimental certification of two entangled pairs after storage. Each correlator $\C_{xy}$ is measured as described in the main text and used to compute the parameter $\mathcal{T}$ of Eq.~(\ref{eq:T_4}). $\tilde{\mathcal{T}}$ is a model-based estimation of the expected $\mathcal{T}$ value in our experiment (see text). There is a good agreement between the two. These results above the bound~(\ref{eq:T_ineq}) of 3.535 certify entanglement for each photon pair released from the QM. The uncertainties represent one standard deviation assuming Poisson statistics for the counts.}
  \label{tbl:sum}
\end{table}

With the recorded four-fold events in hand, we use the entanglement witness to show that the two pairs used to probe the multimode properties of the memory are polarization entangled. Each correlator of Eq.~(\ref{eq:T_4}) is measured for 900 seconds and the sequence is repeated many times (see Table~\ref{tbl:sum}). The experimental value of the entanglement witness is $\mathcal{T} = 3.67\pm 0.06$, two standard deviations above the upper bound~(\ref{eq:T_ineq}) of $3.535$ attainable when only one pair is entangled while the other is separable. One can find all the counts to reconstruct the witness in the~\suppmat.

To understand what limits our measured value of~$\mathcal{T}$, we developed a simple model to predict it using only the measurement of the Bell--CHSH parameter $S$ for a single entangled pair (see the details in the \suppmat). For this we assume that we are measuring two independent pairs and a total quantum state of the form $\rho(\mathcal{V}) = \rho_W(\mathcal{V}) \otimes \rho_W(\mathcal{V})$, where $\rho_W(\mathcal{V}) = \mathcal{V} \ket{\phi_+}\bra{\phi_+} + (1-\mathcal{V}) \openone / 4$ is a two-qubit Werner state with visibility $\mathcal{V}$. We should measure the value $S=2 \sqrt{2}\mathcal{V}$ for a single entangled pair in the state $\rho_W(\mathcal{V})$, and we can use this to calculate the expected value of~$\tilde{\mathcal{T}}$. With the photons retrieved from the QM, we found $S=2.58\pm 0.02$ corresponding to a visibility of $\mathcal{V}=0.912\pm 0.007$, which leads to an expected value of $\tilde{\mathcal{T}}=3.64\pm 0.02$, in agreement with experimentally measured value of $\mathcal{T}$. We note that in this model, a minimum visibility of $\mathcal{V} \simeq 0.85$ (for each identical pair) is required to certify more than one entangled pair, which is more stringent than the case where all measurement outcomes are accessible. 
The maximum value reachable with a state of the form $\rho_W$ is $\mathcal{T} = 8\sqrt{2}/3 \simeq 3.77$ with~$\mathcal{V}=1$. The visibility $\mathcal{V}$ in our experiment was limited equally by the imperfect preparation of the polarization entangled state and multi-pair generation from the SPDC source~\citep{Clausen2014a}.

In conclusion, we quantified the temporal multimode capacity of our solid-state QM using two entangled photon pairs as a probe. To ascertain the high-fidelity storage, we developed an entanglement certification method that can also be used as a Schmidt number witness. The latter does not require any assumptions on the quantum state and works even with a limited set of projective measurements. Our approach can be adapted to certify entanglement involving multiple stored excitations in multiplexing quantum memories harnessing other degrees of freedom such as frequency and spatial modes of light. 

Progress towards a quantum repeater requires the use of a quantum memory that can retrieve photons on-demand using a complete long-duration AFC spin-wave storage \cite{Jobez2015,Gundogan2015,Laplane2016a} with high multimode capacity \cite{Jobez2016}. Alternatively, a scheme based on spectral multiplexing which does not require temporal on-demand readout could be used \cite{Sinclair2014}. Our experiment demonstrating storage of several entangled excitations in ten different temporal modes of a quantum memory together with previous demonstrations open promising perspectives in the direction of long-distance quantum communication.

\section*{Acknowledgements}

We thank Marc-Olivier Renou and Marc Maetz for useful discussions, Boris Korzh for help with the detectors, Alban Ferrier and Philippe Goldner for the crystals and Harald Herrmann and Christine Silberhorn for lending one of the nonlinear waveguides.

\section*{Funding Information}

This work was financially supported by the European Research Council (ERC-AG MEC) and the Swiss National Science Foundation (SNSF). J.~L.~was supported by the Natural Sciences and Engineering Research Council of Canada (NSERC). N.~B acknowledges Swiss National Science Foundation (grant PP00P2-138917 and Starting grant DIAQ). M.~H would like to acknowledge funding from the Swiss  National Science  Foundation (AMBIZIONE Z00P2-161351) and the Austrian Science Fund (FWF) through the START project Y879-N27.

\bibliography{arxiv_submission.bbl}
\newpage

\section*{Appendix for ``Temporal multimode storage of entangled photon pairs''}

\section{Indirect entanglement witnesses}
Here we present the details of the derivation of the entanglement witness used in the main text. Although we expect the final state to be a product of two highly entangled states, we obviously do not want to make this assumption in the derivation of its entanglement certification. For that purpose we treat the underlying events as originating from a $4\times4$ dimensional Hilbert space without any assumption in its internal structure. In the ideal case that state should correspond to a tensor product of two Bell states and thus have a Schmidt number of $4$. If only one of the pairs could retain its entanglement through the storage in the quantum memory its Schmidt number would be at most $2$. If, on the other hand all the entanglement had been destroyed, the resulting state would be completely separable (i.e. Schmidt number $1$).

Ideally we could now derive a Schmidt number witness $W_k$ \cite{Sanpera2001a}, tailored towards this state. One could then decompose this witness into a linear combination of local measurements and acquire all the required coincidences until that witness is constructed and assumption-free entanglement of both pairs is certified. As pointed out in the main text, however, the dead times between the two signals prohibit the acquisition of complete counts in any basis. It is thus impossible to follow the canonical approach of obtaining density matrix elements through local projective coincidences as:
\begin{align}
\langle\alpha,\beta|\rho|\alpha,\beta\rangle=\frac{N_{\alpha,\beta}}{\sum_i\sum_j N_{i,j}:=N_{tot}}\,,
\end{align} 
since $N_{tot}$ inevitably contains terms that cannot be measured in the current setup. What we can measure instead are proportionality constants $\mathcal{T}$ between different count rates, i.e.
\begin{align}
\frac{\sum_{\{a,b\}\in X}c_{a,b}N_{a,b}}{\sum_{\{a,b\}\in Y}d_{a,b}N_{a,b}}=\mathcal{T}\,,
\end{align}
with arbitrary real coefficients $c_{a,b}$ and $d_{a,b}$. Even though we do not have access to $N_{tot}$ we can now use it to rewrite the above as a fraction of corresponding density matrix elements.
\begin{align}
\mathcal{T}=\frac{\frac{1}{N_{tot}}\sum_{\{a,b\}\in X}N_{a,b}}{\frac{1}{N_{tot}}\sum_{\{a,b\}\in Y}N_{a,b}}=\frac{\sum_{\{a,b\}\in X}\langle a,b|\rho|a,b\rangle}{\sum_{\{a,b\}\in Y}\langle a,b|\rho|a,b\rangle}\,.
\end{align}
Through the linearity of the trace we can conclude that the operator
\begin{align}
\mathcal{W}(\mathcal{T})=\sum_{\{a,b\}\in X}|a,b\rangle\langle a,b|-\mathcal{T}\sum_{\{a,b\}\in Y}|a,b\rangle\langle a,b|\,,
\end{align}
has expectation value zero, i.e. $\text{Tr}(\mathcal{W}(\mathcal{T})\rho)=0$. Now we can continue to look at the following optimization problem
\begin{align}
\min_{\sigma\in SEP}\text{Tr}(\mathcal{W}(\mathcal{T})\sigma):=E_{min}\,.
\end{align}
If we can prove that $E_{min}>0$, we have thus proven entanglement of the state $\rho$, through the induced witness $\mathcal{W}(\mathcal{T})$. Even though the above optimization problem is convex, it may nonetheless be hard to solve. We can however use the concept of positive maps to address it through a relaxation (even in the multipartite case, see \cite{Lancien2015a}). As an exemplary case consider the partial transpose map, i.e.
\begin{align}
\min_{\sigma\in PPT}\text{Tr}(\mathcal{W}(\mathcal{T})\sigma):=E_{PPT}\leq E_{min}\,.
\end{align}
Since the explicit form of the above optimization problem is a semi-definite program (SDP), it can be easily solved using SDP solvers, where the maximization of the dual yields analytic lower bounds on the minimum. It is even possible to address the dimensionality of entanglement in this context. Since the negativity $\mathcal{N}(\rho)$ is bounded for states of Schmidt rank $k$ as $\frac{k-1}{2}$ \cite{Eltschka2013a}, we can make use of its variational form $\mathcal{N}(\rho)=\min(\text{Tr}(M_{-}))$,  s.t. $\rho^{T_A}=M_{+}-M_{-}, M_{\pm}\geq 0$ to introduce the following SDP
\begin{align}
\min_{\sigma}\text{Tr}(\mathcal{W}(\mathcal{T})\sigma):=E_{k}, \\\text{s.t. } \sigma^{T_A}=M_{+}-M_{-}, M_{\pm}\geq 0, \text{Tr}(M_{-})\leq \frac{k-1}{2}\,.
\end{align}
This is again an SDP, and thus possible to compute analytic bounds. Now it is obvious that if $E_k>0$ the Schmidt number (i.e. entanglement dimensionality) of the experimental state $\rho$ has to be greater than $k$.

Now applying the above formalism to the two-photon pair experiment we can use
\begin{equation}
\mathcal{T} =   \frac{1}{N}( C_{00}+C_{01}+C_{10}-C_{11})   
\label{eq:T_4_sm}
\end{equation}
where we have defined correlation functions 
\begin{equation}
C_{xy}=  \sum_{a,b = \pm 1} ab  \,  N_{ab,\bar{a}\bar{b}|xy}
\label{eq:e_sm}
\end{equation}
and a normalization factor 
\begin{equation}
N = \frac{1}{4}\sum_{x,y=0,1} \sum_{a,b=\pm1}  N_{ab,\bar{a}\bar{b}|xy}.
\end{equation}
Now we can optimize over all $T$ such that:
\begin{align}
\mathcal{T}_{PPT}:= \max \mathcal{T}\\
\text{s.t. } E_{PPT}\leq 0
\end{align}
and
\begin{align}
\mathcal{T}_{1-photon}:= \max \mathcal{T}\\
\text{s.t. } E_{2}\leq 0
\end{align}
In the last program we can furthermore add Schmidt number witnesses \cite{Sanpera2001a} to be non-violated as a linear constraint.
We have implemented these programs with MATLAB using the SDPT3 solver and the YALMIP package \cite{yalmip2004} and found that $\mathcal{T}_{PPT}=2\sqrt{2}$ and with a bounded negativity and choosing the optimal Schmidt number witness for the unconstrained state the value for one photon is $\mathcal{T}\approx 3.535$, thus showing that the induced entanglement witnesses $\mathcal{W}(\mathcal{T})$ indeed proves that more than one entangled pair is required to explain the observed correlations.

\section{Consistency check}

The consistency of the experimentally obtained value of $\mathcal{T}$ can be verified with the following model. Assume two independent photon pairs, each pair being in a Werner state with visibility $\mathcal{V}$: $\rho_W(\mathcal{V})  =  \mathcal{V}  \ket{\phi_+} \bra{\phi_+} +(1-\mathcal{V}) \openone / 4 $, where $\ket{\phi_+} = \frac{1}{\sqrt{2}} (\ket{00}+ \ket{11}) $ is a Bell state. Hence the total state is of the form $
\rho(\mathcal{V}) = \rho_W(\mathcal{V}) \otimes \rho_W(\mathcal{V})$. For such a state, the witness value $\mathcal{T}$ can be directly expressed in function of the visibility
\begin{equation}
\mathcal{T} = \frac{4 \sqrt{2} \mathcal{V}}{1 + \mathcal{V}^2 / 2}.
\end{equation}
Equivalently, this can be expressed in terms of the Bell-CHSH parameter $S$ for a single photon pair:
\begin{equation}
\mathcal{T} = \frac{2S}{1 + S^2 / 16}.
\end{equation}
Here the CHSH parameter is given by 
\begin{equation}
S = \sum_{x,y=0,1} (-1)^{xy} E_{xy}
\end{equation}
where we have defined the usual correlators $E_{xy} = \sum_{a,b=\pm1 } ab p(ab|xy)$. For a single photon pair in a Werner state $\rho_W(\mathcal{V})$, the CHSH value is a simple function of the visibility: $S = 2\sqrt{2} \mathcal{V}$. Notice that the measurement settings used in our witness are the optimal settings for testing CHSH on the Bell state $\ket{\phi_+}$. 

Experimentally, we measured for a single photon pair $S = 2.58 \pm 0.02$. Given the above model, we thus expect a witness value of $\tilde{\mathcal{T}} = 3.64 \pm 0.02$, in excellent agreement with our data.

\begin{table}[t!]
\centering
\renewcommand{\arraystretch}{2}
\begin{ruledtabular}
\begin{tabular}{c|cccc|cccc}
     & \multicolumn{4}{c|}{Stored}       & \multicolumn{4}{c}{Transmitted}  \\ \hline
            &$x_0,y_0$  &  $x_0,y_1$  &  $x_1,y_0$  &  $x_1,y_1$  & $x_0,y_0$  &  $x_0,y_1$  &  $x_1,y_0$  &  $x_1,y_1$ \\ \hline
$N_{++,--}$ & 76     & 70     & 61    & 4     & 195    & 220    & 200    & 6 \\
$N_{--,++}$ & 112    & 113    & 112   & 6   & 216    & 227    & 198   & 11    \\
$N_{+-,-+}$ & 1      & 2      &  5    & 92      & 10     & 12     & 11   & 222 \\
$N_{-+,+-}$ & 2      & 7      & 4   & 82     & 9     & 10     & 13    & 223   
\end{tabular}
\end{ruledtabular}
\caption{
Experimental certification of two entangled pairs. The total number of four-fold events detected for each combination of four detectors (two detectors on each side of the experiment). The data was collected for the maximum delay between photon pairs of 50~ns. For this reason only 4 detector combinations are available due to their deadtime. Each column is used to calculate the correlators $\C_{xy}$ (Eq.~(\ref{eq:e_sm})) for different sets of projective measurements as described in the main text. Their values are used to compute the parameter $T$ of Eq.~(\ref{eq:T_4_sm}) for stored entangled photon pairs or transmitted through the QM. For stored photons we obtained the value of entanglement witness of $\mathcal{T} = 3.67(6)$, while for the transmitted part it is $3.63(4)$. Both values are above the entanglement bound to certify two entangled pairs.
}
  \label{tbl:sum2}
\end{table}

\begin{figure}[h!]
\centering
\includegraphics[width=0.45\textwidth]{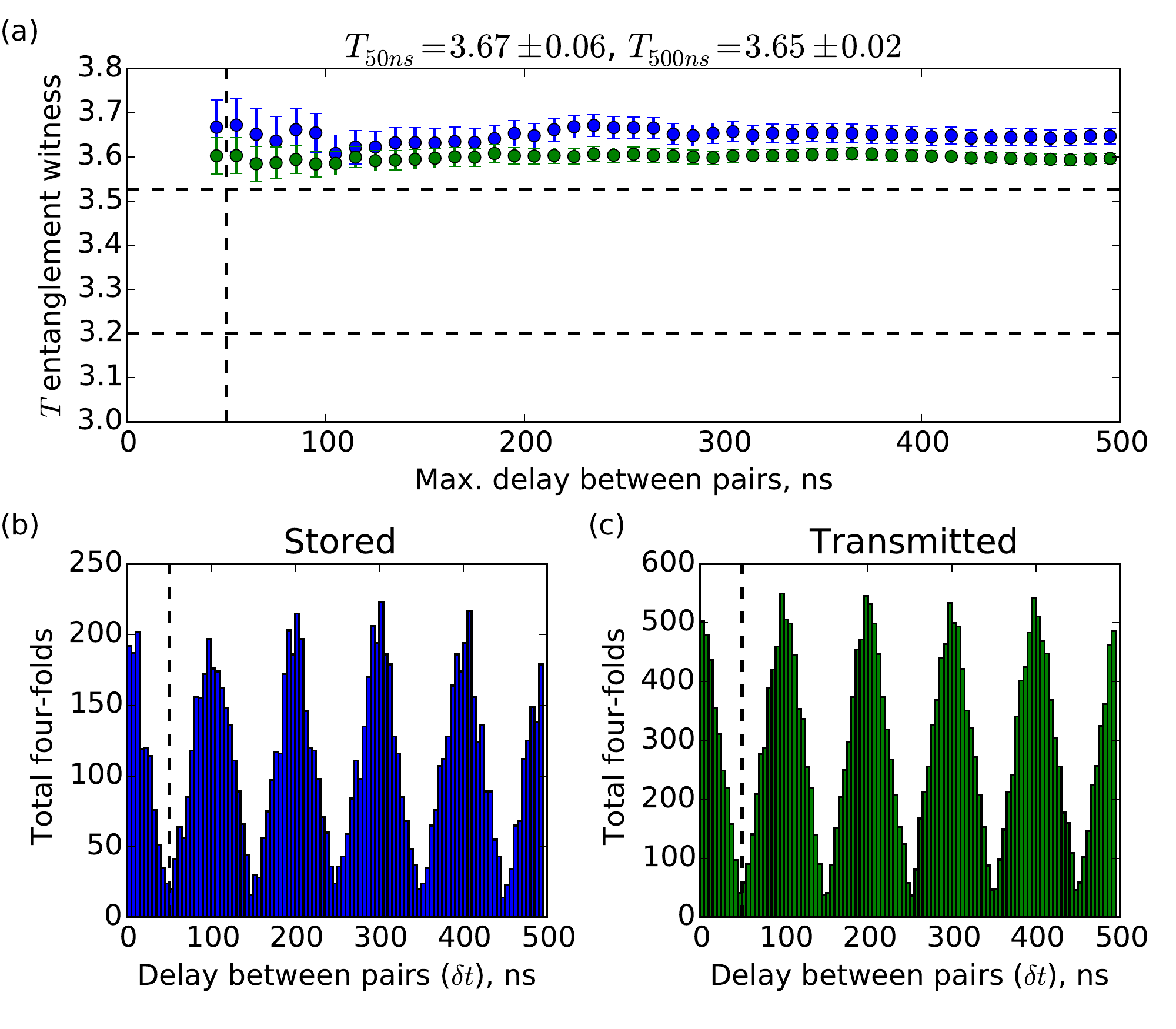}
\caption{(a) The value of entanglement witness as a function of the delay between photon pairs which were transmitted or stored in the quantum memory. Bigger delay decreases standard deviation due to the bigger statistics. All the values are above entanglement bound (dash line), which certifies the presence of entanglement for both photon pairs. The values for 50~ns and 500~ns maximum delay are equal inside standard deviation. (b) The total number of quadruples  is plotted in function of the delay between photon pairs which were transmitted (right graph) or stored in the quantum memory (left). Triangular shape comes from the square shape of the pump and convolution. To certify storage of both photon pairs we consider a maximum delay of 50~ns between two photon pairs. For longer delay, one cannot certify that two stored for at least some time.  The number of stored quadruples from the first pulse or any other is the same which underlines that the efficiency of the QM doesn't depend on the number of stored excitations.
}
\label{fig:results}
\end{figure}

\begin{figure*}[htp]
\subfloat[]{%
  \includegraphics[clip,width=\textwidth]{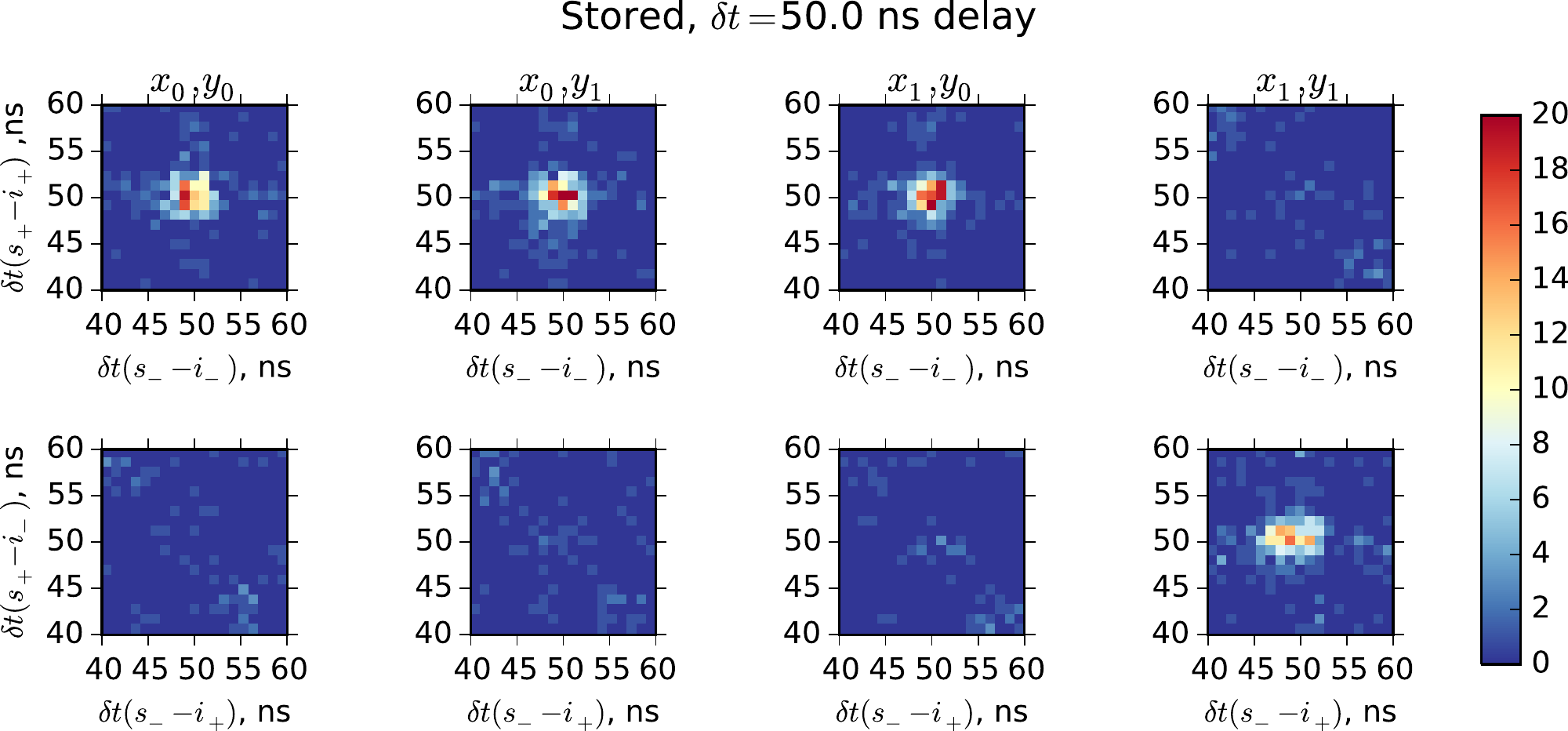}%
}

\subfloat[]{%
  \includegraphics[clip,width=\textwidth]{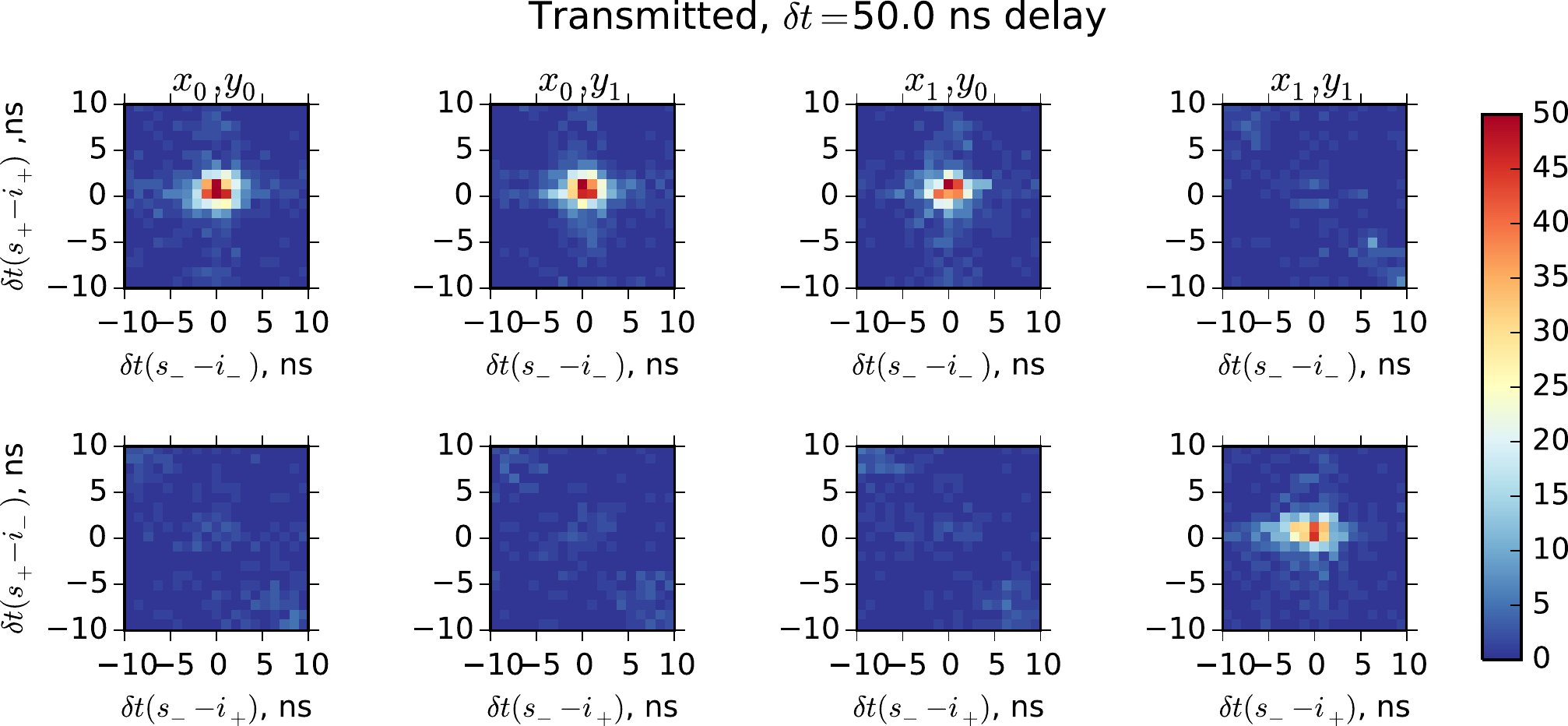}%
}
\caption{ Histograms of the number of four-fold coincidences with the stored (a) and transmitted (b) signal photon that were used to verify entanglement for both photon pairs.  They are shown as a function of delays $\delta t(s_k - i_l)$ between detector $D^{(s)}_k$  of the signal analyzer ($k =  +; -$) and detector $D^{(i)}_l$ ($l =  +; -$), where $D^{(i)}$ is one of the two SNSPDs used to detect idler photon. The histograms were measured in four different bases that are equivalent to the CHSH inequality violation. The binning of the histogram in each direction was set to be 0.486~ps. The coincidence window of 5~ns around the center of the histograms was used to collect the four fold events and calculate entanglement witness~(\ref{eq:T_4_sm}). 
}
\end{figure*}

\newpage
\noindent

\newpage
\noindent

\newpage
\noindent

\begin{figure*}[!t]
\centering
\subfloat[]{
	\def\svgwidth{0.45\textwidth}
	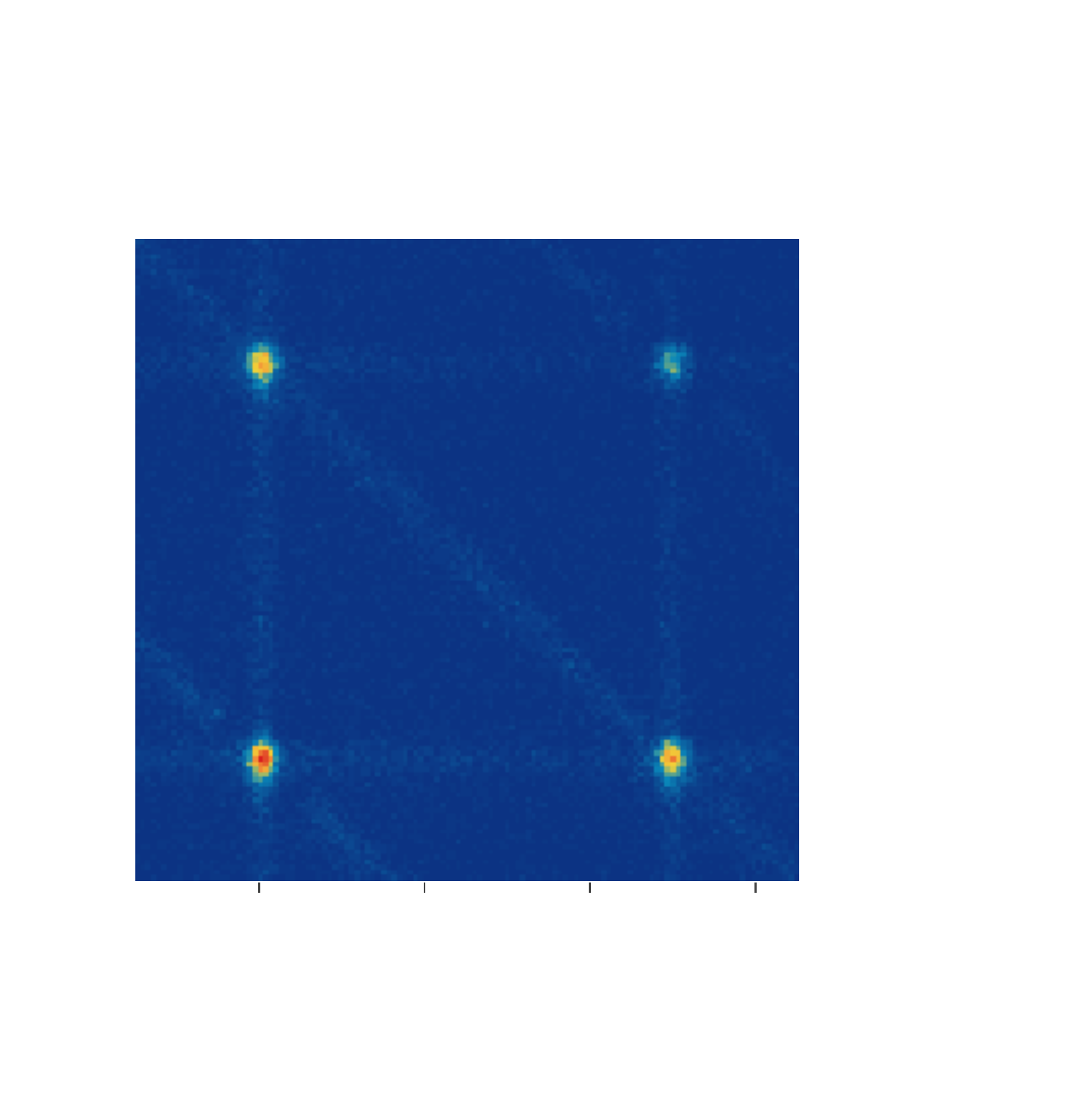
}
\subfloat[]{
	\def\svgwidth{0.45\textwidth}
	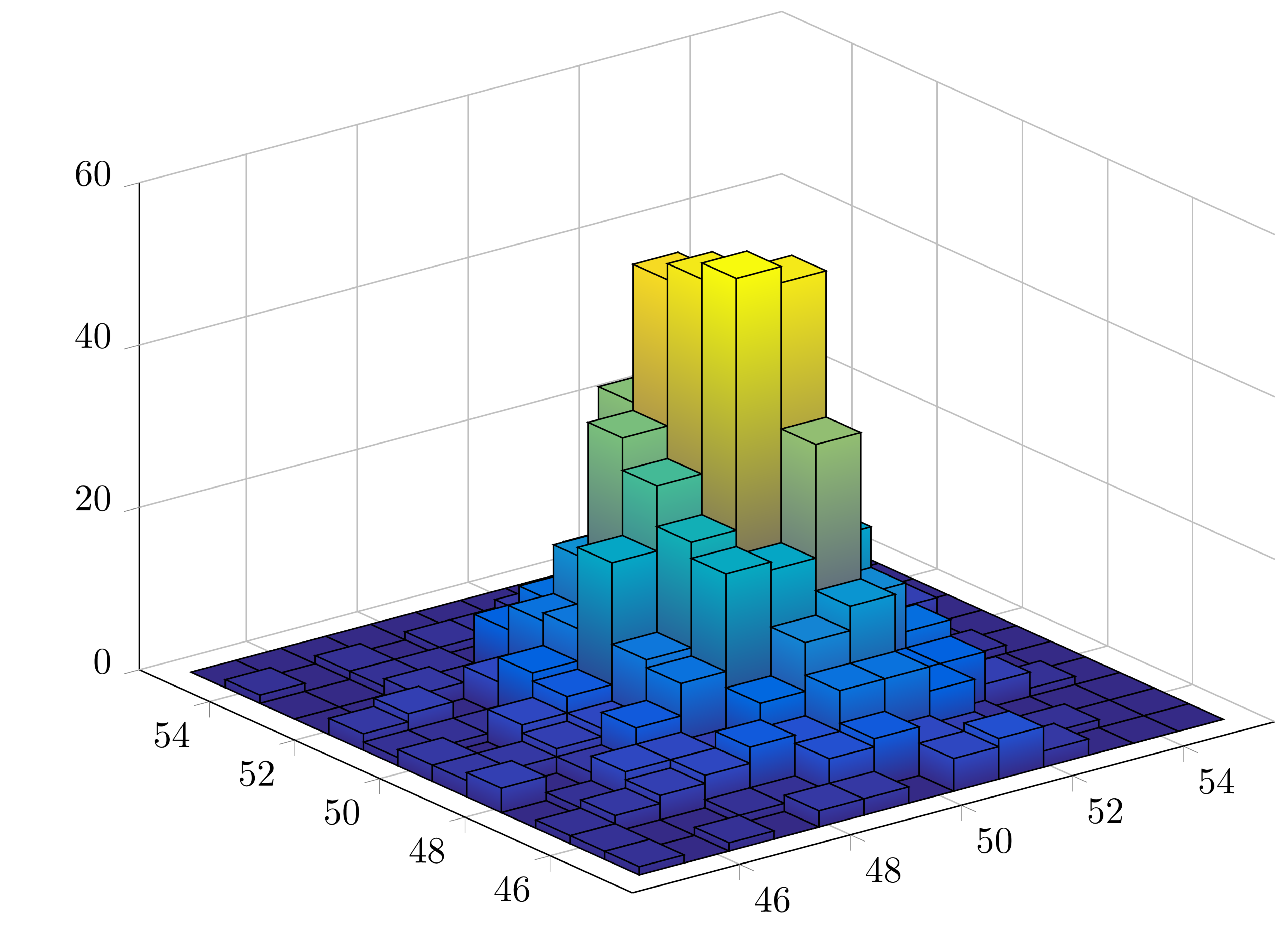
}
\caption{(a) The total number of quadruple events plotted as a function of the delays between different pairs of detectors by which they were detected. The maximum delay of 500~ns was considered. Four different coincidence peaks correspond to four different cases: (A) both signal photons were stored in the QM, (C) both signal photons were transmitted without absorption in the crystal and two cases (B, D) where only one of the signal photon was stored while other was transmitted. (b) The total number of stored quadruples (case A) as a function of the delays between different pairs of detectors.
}
\label{fig:3dhist}
\end{figure*}

\end{document}